\documentclass{article}
\usepackage{spconf,amsmath,graphicx,hyperref}
\usepackage{enumitem}
\usepackage{amsmath}
\usepackage{amssymb}
\usepackage{booktabs}
\usepackage{array}

\title{Beyond Visible Speakers: Context-Aware Active Speaker Detection with Speaker Registration}
\name{Pu Wang~$^{1}$ \qquad Hugo Van hamme~$^{2}$}
  
  \address{$^{1, 2}$~KU Leuven, Department of Electrical Engineering, Leuven, Belgium \\
      $^{1}$~pu.wang@esat.kuleuven.be, $^{2}$~hugo.vanhamme@kuleuven.be}
\begin{document}
\ninept
\maketitle
\begin{abstract}
Active speaker detection (ASD) relies on audiovisual correspondence between speech and visible faces, yet this correspondence becomes unreliable in crowded scenes, or when visual evidence is unavailable. We introduce ROAM-ASD-C and ROAM-ASD-CR to extend ROAM-ASD with contextual and identity-aware reasoning. ROAM-ASD-C jointly reasons across simultaneously visible candidates, allowing other faces to provide complementary evidence when the target observation is ambiguous or incomplete. ROAM-ASD-CR incorporates registered-speaker representations derived from enrollment speech, allowing detection of whether a speaker is active even when that speaker is not visually observed. ROAM-ASD-C achieves state-of-the-art performance across five ASD benchmarks: $96.8\%$ mAP on AVA, $98.9\%$ on WASD, $88.8\%$ on UniTalk, $98.7\%$ on Talkies, and $99.5\%$ on ASW. It also improves ROAM-ASD by up to $13.3$ mAP under $75\%$ target-track visual removal. 
\end{abstract}
\section{Introduction}
\label{sec:intro}
Active speaker detection (ASD) aims to determine whether a visible person is speaking in an audiovisual scene. It is an important component of applications such as speaker diarization and multimodal human-computer interaction. Recent ASD systems, such as ASC~\cite{alcazar2020active}, MAAS~\cite{alcazar2021maas}, ASDNet~\cite{kopuklu2021design}, Light-ASD~\cite{liao2023light}, LoCoNet~\cite{wang2024loconet}, TalkNCE~\cite{jung2024talknce}, TS-Talk~\cite{jiang23c_interspeech}, and MuSED~\cite{tao2024enhancing}, have achieved strong performance on conventional benchmarks, with performance on AVA~\cite{roth2020ava} approaching saturation at over $95\%$ mean average precision (mAP). Nevertheless, ASD remains more challenging in unconstrained recordings, where multiple candidate speakers, interfering voices, and domain variation make the correspondence between acoustic speech and visible faces ambiguous. This gap is reflected in recent in-the-wild benchmarks such as WASD~\cite{roxo2024wasd} and UniTalk~\cite{nguyen2025revisiting}, where the best reported baseline performance reaches only $93.7\%$ and $83.2\%$ mAP, respectively.

The core of an ASD system is audio-visual fusion. Most existing methods rely primarily on global full-face features as visual evidence, which can be distracted by non-speech facial activity under in-the-wild conditions~\cite{roxo2024wasd}. Many systems adopt predefined pairwise fusion paths, such as cross-attention, which can be less robust when one modality becomes unreliable or incomplete, for example under face occlusion. In our previous work, ROAM-ASD~\cite{wang2026roam}, we addressed these two limitations for challenging in-the-wild conditions. ROAM-ASD augments global face representations with fine-grained localized visual cues, such as mouth information, and fuses audio, face, and mouth streams through flexible joint self-attention with modality-dropout training. This design improves cross-domain generalization and robustness to missing observations, raising the state of the art (SOTA) on WASD and UniTalk to $98.8\%$ and $87.9\%$ mAP, respectively.

However, ROAM-ASD still makes its decision independently for each visible face and therefore does not exploit relationships among simultaneously visible speaker candidates. Such context can be particularly useful in crowded or ambiguous scenes, where multiple faces may exhibit speech-like movements or acoustic interference from background noise and overlapping speech may be present. Prior works have demonstrated the benefit of this form of reasoning. ASC~\cite{alcazar2020active} models relations among multiple visible speakers and their temporal context. MAAS~\cite{alcazar2021maas} formulates speaker assignment through multimodal graph reasoning. LoCoNet~\cite{wang2024loconet} further incorporates inter-speaker interactions together with long-term within-speaker context. Motivated by this, we extend ROAM-ASD with inter-face contextual reasoning among simultaneously visible candidates. 

Despite these benefits, such contextual reasoning remains restricted to the \emph{visible} candidate set. When visual observations are unavailable altogether, conventional face-level ASD loses the audio-visual correspondence required to assign speech to a \emph{particular} person. Since the same scene-level audio is available to every candidate, speaker identity becomes ambiguous without an additional identity cue. A complementary direction is to incorporate speaker identity into audio-visual modeling, for example through target-speaker detection~\cite{jiang23c_interspeech}. Such approaches show that prior speaker information can help disambiguate speech activity. Motivated by this, we introduce \emph{speaker registration}, allowing enrolled speaker identities to be represented in addition to visible face candidates.
\begin{figure*}[!htbp]
\centerline{\includegraphics[width=1.0\linewidth]{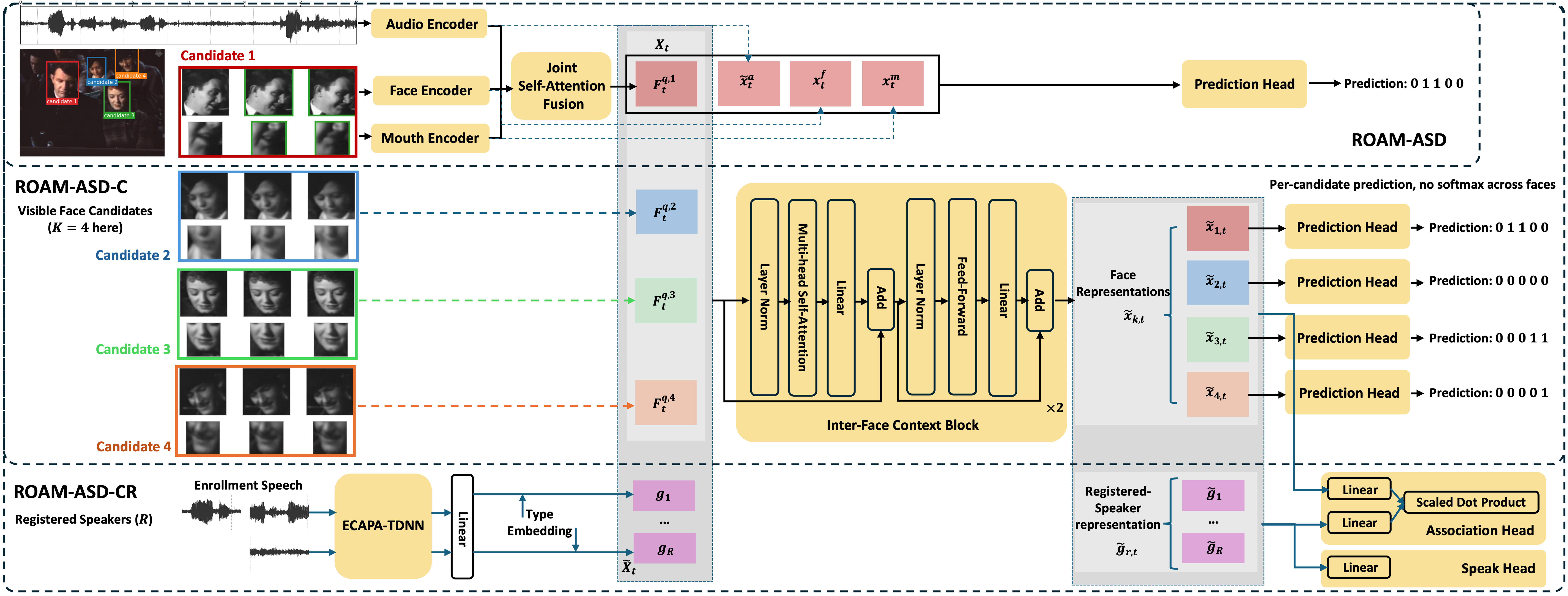}}
\vspace{-1.2em}
\caption{Overview of ROAM-ASD~\cite{wang2026roam}, ROAM-ASD-C, and
ROAM-ASD-CR. ROAM-ASD-C adds inter-face contextual reasoning across
temporally aligned face candidates, while ROAM-ASD-CR further
incorporates enrollment speech as registration tokens for
registered-speaker activity and face association. Original ROAM-ASD
skip connections are omitted for clarity but retained unchanged.}
\label{fig:architecture}
\vspace{-4mm}
\end{figure*}

In this work, we extend ROAM-ASD in two complementary directions.  First, we introduce ROAM-ASD-C, which augments the track-level ROAM-ASD representation with inter-face contextual reasoning across simultaneously visible candidates. Candidate representations are jointly processed before the final per-face activity predictions, allowing other visible candidates to provide complementary evidence when several candidates compete for the same acoustic evidence or when the target candidate's own observations are ambiguous or incomplete. The final ASD decisions remain independent binary predictions, preserving the ability to represent silence and simultaneous speakers. Second, we introduce ROAM-ASD-CR, which further augments the candidate set with registered-speaker representations derived from enrollment speech. Registered speakers participate in the context block alongside visible face candidates and receive their own activity predictions. This extends ASD beyond the visible speakers by allowing the model to determine whether a registered speaker is speaking even when visual observations are entirely unavailable. When visual evidence is available, the registered identity can additionally be associated with the corresponding face. We conduct experiments across five ASD benchmarks to evaluate inter-face contextual reasoning, robustness to missing target-track observations, and speaker registration. ROAM-ASD-C consistently improves ROAM-ASD across all five benchmarks, with larger gains in crowded, difficult, and visually degraded conditions, and when the target candidate's own observations are incomplete. ROAM-ASD-CR further preserves standard face-level ASD performance while enabling registered-speaker activity detection when visual observations are entirely unavailable. The main contributions are:
\begin{itemize}[itemsep=0pt, topsep=0pt, leftmargin=*]
\item We introduce ROAM-ASD-C, which extends ROAM-ASD from independent face-track prediction to inter-face contextual reasoning among simultaneously visible candidates.
\item ROAM-ASD-C consistently improves ROAM-ASD across five ASD benchmarks, achieving new SOTA performance of $96.8\%$, $98.9\%$,
$88.8\%$, $98.7\%$, and $99.5\%$ mAP on AVA, WASD, UniTalk, Talkies, and ASW, respectively. The gains become larger under more challenging and crowded conditions. Examples are available at\url{https://wangpuup.github.io/ROAM-ASD-C/crowds.html}.
\item ROAM-ASD-C improves robustness to incomplete target-track observations. Under $75\%$ visual evidence removal, it improves over ROAM-ASD by $3.1$, $8.8$, $13.3$, and $10.7$ mAP on AVA, WASD, UniTalk, and ASW, respectively. Examples are available at \url{https://wangpuup.github.io/ROAM-ASD-C/modality.html}.
\item We introduce ROAM-ASD-CR, which extends ASD beyond the question ``which visible person is speaking?'' to ``is this registered person speaking, and which visible face corresponds to that identity?'' Speaker registration provides the identity cue that conventional face-level ASD lacks when visual correspondence is unavailable, enabling ASD beyond the currently visible scene.
\end{itemize}

\section{Model}
\label{sec:model}
We extend ROAM-ASD~\cite{wang2026roam} from independent track-level prediction to scene-level reasoning. Specifically, we introduce (1) an \emph{inter-face context block} that enables information exchange among simultaneously visible speaker candidates, yielding ROAM-ASD-C, and (2) \emph{speaker registration}, which augments the candidate set with enrolled voice representations, yielding ROAM-ASD-CR. 
The first extension allows ambiguous or incomplete observations of one visible candidate to be interpreted in the context of the other people in the scene, while the second extends the reasoning space beyond the currently visible candidate set. We first briefly recap ROAM-ASD.

ROAM-ASD~\cite{wang2026roam} performs track-level ASD using audio, full-face, and mouth-region inputs. Separate encoders produce audio, face, and mouth representations, which are fused using joint self-attention with modality-agnostic query tokens. The resulting fused query representations $\mathbf{F}^{q}\in\mathbb{R}^{T_v\times d}$ provide one $d$-dimensional representation for each of the $T_v$ video frames. Modality dropout during training allows it to operate under variable input availability. In this work, we retain the original encoders, multimodal fusion, modality-dropout strategy, and per-face prediction head, and introduce contextual reasoning on the fused query representations before prediction.

\textbf{Inter-face contextual reasoning (ROAM-ASD-C).}
ROAM-ASD processes each face track independently. This can be limiting in crowded scenes, where the active-speaker decision for one candidate may benefit from evidence provided by other simultaneously visible candidates. We therefore introduce an inter-face context block that performs joint reasoning across visible people. To construct the candidate set for contextual reasoning, we first group temporally overlapping face tracks into a common scene. Each track is treated as one speaker candidate. Because tracks may have different frame timestamps, their fused query representations are temporally aligned before being passed to the context block. We use the timestamps of the longest track as the reference grid, and at each time step $t$, every candidate $k$ contributes the representation \(\mathbf{F}^{q,k}\) nearest to $t$ within half a frame step, otherwise, that candidate is masked from the subsequent context block. The aligned representations are stacked along the candidate dimension,
$\mathbf{X}_{t} = \left[\mathbf{F}^{q,1}_{t};\ldots;\mathbf{F}^{q,K}_{t}\right]\in\mathbb{R}^{K\times d}$ ($K\leq6$ and $d=256$), as illustrated in Fig.~\ref{fig:architecture}. Scenes containing more than six overlapping tracks are split into groups of at most six candidates.

The inter-face context block consists of two pre-norm Transformer layers with 256-dimensional, 4-head self-attention and a 512-dimensional feed-forward network with GELU activation. Self-attention is performed over the candidate dimension, i.e., the $K$ rows of $\mathbf{X}_t$, independently at each time step, allowing the visible face candidates to exchange information directly. Each self-attention and feed-forward sub-layer is followed by an additional zero-initialized linear projection. These projections make the context block an identity mapping at initialization, so the original per-face
representations are initially preserved while cross-candidate interactions are learned gradually. Groups containing a single face candidate bypass it. After the context block, each of the $K$ output representations corresponds to one face candidate enriched with information from the other visible candidates. These contextualized fused query representations are returned to their corresponding face tracks and passed to the unchanged ROAM-ASD prediction head together with the original skip features. The head is applied to each candidate separately, with no softmax across faces, allowing both silence and simultaneous speakers to be represented.

\textbf{Speaker registration (ROAM-ASD-CR).}
ROAM-ASD-C enables contextual reasoning among the \emph{visible} face candidates in a scene and cannot represent a known speaker when that speaker is not visually observable. To extend contextual reasoning beyond the visible candidates, ROAM-ASD-CR introduces \emph{speaker registration}, in which enrollment speech is encoded as an additional speaker candidate that participates in the same context block as shown in Fig.~\ref{fig:architecture}.

To construct this representation for registered speaker $r$, a frozen ECAPA-TDNN speaker encoder extracts 192-dimensional embeddings from 1-3 enrollment segments of at most $3$~s. Each embedding is $\ell_2$-normalized, and their mean is re-normalized to obtain the enrollment representation $\mathbf{s}_r$. $\mathbf{s}_r$ is projected to the model dimension and combined with a learned registration-type embedding to form a registration token $\mathbf{g}_r\in\mathbb{R}^{d}$.
The token is time invariant and is included at every time step alongside the visible face candidates: $\widetilde{\mathbf{X}}_{t}=\left[\mathbf{X}_{t};\mathbf{g}_1;\ldots;\mathbf{g}_R\right]$, where $R$ is the number of registered speakers, as illustrated in Fig.~\ref{fig:architecture}. The augmented candidate set is processed by the context block from ROAM-ASD-C, allowing the registered-speaker and face representations to exchange information directly. The context block produces one contextualized representation for each face candidate and each registered speaker. We denote these representations by $\widetilde{\mathbf{x}}_{k,t}$ and
$\widetilde{\mathbf{g}}_{r,t}$, respectively. At most six registration slots are used during training. When $R=0$, no registration tokens are appended, and the model reduces to ROAM-ASD-C.

After contextual reasoning, the contextualized face representations are passed to the unchanged ROAM-ASD prediction head for per-face ASD. Each registered-speaker representation is used to answer two complementary questions: whether the registered speaker is currently speaking and, if so, which visible face corresponds to that speaker. A linear \emph{speak head} maps $\widetilde{\mathbf{g}}_{r,t}$ to a logit $o_{r,t}$ indicating whether registered speaker $r$ is speaking. An \emph{association head} linearly projects the registered-speaker and face representations into a shared space and computes their scaled dot-product similarity $a_{r,k,t}
=
\frac{
\left(\mathbf{W}_{r}\widetilde{\mathbf{g}}_{r,t}\right)^{\top}
\left(\mathbf{W}_{f}\widetilde{\mathbf{x}}_{k,t}\right)
}{\sqrt{d}}$, where $\mathbf{W}_{r}$ and $\mathbf{W}_{f}$ are learned projections for the registration and face representations, respectively. The resulting score $a_{r,k,t}$ measures the association between registered speaker $r$ and visible face candidate $k$ at time $t$.

The registration heads are trained through self-enrollment without additional speaker-identity annotation. A speaker appearing in the scene is treated as a \emph{matched registrant}, whose corresponding face track provides the speaking and association targets. A speaker enrolled from another video but not present in the current scene is treated as an \emph{out-of-scene registrant} and provides negative supervision for speaking activity. Based on this, we introduce two registration-specific losses in addition to the ROAM-ASD objective.

The speak head is trained to determine whether each registered speaker is currently speaking. Let $\mathcal{T}_r$ denote the valid supervised time steps for registrant $r$. When enrollment speech overlaps the current training window, the corresponding interval is excluded from $\mathcal{T}_r$, such that the model is not supervised on the enrollment speech itself. Let $\mathcal{R}$ contain the registrants for which $\mathcal{T}_r\neq\emptyset$. We use binary cross-entropy (BCE), 
\vspace{-0.2em}
\begingroup
\setlength{\abovedisplayskip}{0.5em}
\setlength{\belowdisplayskip}{0.3em}
\setlength{\abovedisplayshortskip}{0pt}
\setlength{\belowdisplayshortskip}{0pt}
\begin{equation}
\mathcal{L}_{\mathrm{spk}}
=
\frac{1}{|\mathcal{R}|}
\sum_{r\in\mathcal{R}}
\frac{1}{|\mathcal{T}_r|}
\sum_{t\in\mathcal{T}_r}
\operatorname{BCE}
\left(
\sigma(o_{r,t}),y_{r,t}
\right),
\end{equation}
\endgroup
where $y_{r,t}=1$ if matched registrant $r$ is speaking and $0$ otherwise; for a out-of-scene registrant, $y_{r,t}=0$.

The association head is trained to associate an active registered speaker with its visible face. Association supervision is therefore applied only when a matched registrant is speaking and its corresponding face is visible. Let $\mathcal{T}^{+}_r\subseteq\mathcal{T}_r$ contain these time steps, let $\mathcal{R}_m$ denote the matched registrants for which $\mathcal{T}^{+}_r\neq\emptyset$, and let $k_r$ denote the face candidate corresponding to registrant $r$. Applying a softmax over the $K$ association scores, the association loss is
\vspace{-0.2em}
\begingroup
\setlength{\abovedisplayskip}{0.5em}
\setlength{\belowdisplayskip}{0.3em}
\setlength{\abovedisplayshortskip}{0pt}
\setlength{\belowdisplayshortskip}{0pt}
\begin{equation}
\mathcal{L}_{\mathrm{assoc}}
=
-\frac{1}{|\mathcal{R}_m|}
\sum_{r\in\mathcal{R}_m}
\frac{1}{|\mathcal{T}^{+}_r|}
\sum_{t\in\mathcal{T}^{+}_r}
\log
\frac{
    \exp(a_{r,k_r,t})
}{
    \sum_{k=1}^{K}\exp(a_{r,k,t})
}.
\end{equation}
\endgroup

The complete training objective combines the original ASD objective with the two registration-specific objectives: $\mathcal{L}
=\mathcal{L}_{\mathrm{asd}}+0.5\,\mathcal{L}_{\mathrm{spk}}+0.25\,\mathcal{L}_{\mathrm{assoc}}$, where $\mathcal{L}_{\mathrm{asd}}$ is the original masked frame-level ROAM-ASD loss, also used for ROAM-ASD-C.

\section{Experiments}
We evaluate ROAM-ASD-C and ROAM-ASD-CR on the same five datasets and follow the same setup as ROAM-ASD~\cite{wang2026roam}: AVA~\cite{roth2020ava}, and four in-the-wild benchmarks, ASW~\cite{kim21k_interspeech}, Talkies~\cite{alcazar2021maas}, WASD~\cite{roxo2024wasd}, and UniTalk~\cite{nguyen2025revisiting}. WASD contains five conditions: \emph{Optimal Conditions (OC)}, \emph{Speech Impairment (SI)}, \emph{Face Occlusion (FO)}, \emph{Human Voice Noise (HVN)}, and \emph{Surveillance Settings (SS)}. UniTalk covers \emph{Underrepresented Languages}, \emph{Background Noise}, \emph{Crowded Scenes}, and \emph{Mixed Conditions}. For ROAM-ASD-CR, registration is disabled for a training scene with probability $0.2$; otherwise, each scene speaker with available speech is independently registered with probability $0.8$. To reflect realistic scenarios in which a set of known speakers is enrolled in advance but only a subset is present in a given scene, we additionally sample zero to four out-of-scene
registrants. We refer to this setting as \emph{over-registration}.

We conduct three sets of experiments. First, we evaluate the effect of inter-face contextual reasoning on face-level ASD, including analyses across different numbers of simultaneously visible faces and different scene conditions. 
Second, we evaluate whether other visible candidates can provide complementary evidence when the target candidate's \emph{own} observations become incomplete. Following the continuous missing-input protocol of ROAM-ASD~\cite{wang2026roam}, we use holes spanning 5\%, 15\%, or 30\% of each target track, repeated to reach 30\% total coverage, as well as single holes spanning 50\% or 75\% of the track and complete (100\%) stream removal. Missing segments are applied to the \emph{target track}, while the other simultaneously visible candidates remain mostly available. We report mAP over labeled frames within the missing intervals, using identical missing segments for ROAM-ASD and ROAM-ASD-C.
Third, we evaluate whether speaker registration can extend ASD beyond the visible candidate set. We remove the face and mouth inputs for \emph{all candidates} over the entire track while retaining scene audio and registration tokens. We report registered-speaker average precision (AP), which measures whether an enrolled speaker is speaking, and association accuracy, which measures whether an active registered speaker is associated with the correct visible face. Under over-registration, we report the false-alarm (FA) rate for
out-of-scene registrants.

\section{Results}
\begin{table}[!htbp]
\vspace{-2.4em}
\centering
\caption{Overall mAP ($\uparrow$, \%) across five benchmarks, compared with the best-performing prior methods and ROAM-ASD.}
\vspace{0.2em}
\label{tab:overall}
\resizebox{0.43\textwidth}{!}{
\setlength{\tabcolsep}{6pt}
\begin{tabular}{lccccc}
\toprule
Method & AVA & WASD & UniTalk & Talkies & ASW \\
\midrule
TalkNCE~\cite{jung2024talknce}
    & -- & -- & $83.2$ & -- & -- \\
Light-ASD~\cite{liao2023light}
    & $94.1$ & $93.7$ & -- & -- & -- \\
MuSED~\cite{tao2024enhancing}
    & $95.6$ & -- & -- & -- & $98.3$ \\
LoCoNet~\cite{wang2024loconet}
    & $95.2$ & -- & $82.2$ & $96.1$ & -- \\
\midrule
ROAM-ASD~\cite{wang2026roam}
    & $96.5$ & $98.8$ & $87.9$ & $98.2$ & $99.3$ \\
\textbf{ROAM-ASD-C}
    & $\mathbf{96.8}$ & $\mathbf{98.9}$ & $\mathbf{88.8}$
    & $\mathbf{98.7}$ & $\mathbf{99.5}$ \\
\textbf{ROAM-ASD-CR}
    & $96.7$ & $\mathbf{98.9}$ & $\mathbf{88.8}$ & -- & $\mathbf{99.5}$ \\
\bottomrule
\end{tabular}
}
\vspace{-1em}
\end{table}
\begin{figure}[!htbp]
\vspace{-1em}
\centerline{\includegraphics[width=0.99\linewidth]{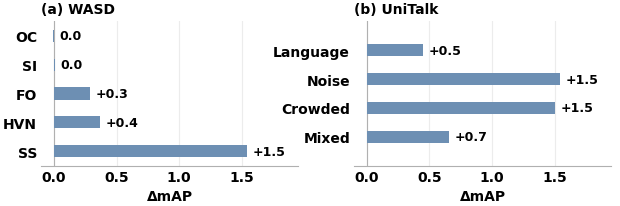}}
\vspace{-1.2em}
\caption{Inter-face context gain ($\Delta\mathrm{mAP} =
\mathrm{mAP}_{\text{ROAM-ASD-C}}-\mathrm{mAP}_{\text{ROAM-ASD}}$)
across (a) WASD and (b) UniTalk categories.}
\label{fig:wasdunitalk}
\vspace{-2.2em}
\end{figure}
\begin{figure}[!htbp]
\centerline{\includegraphics[width=1.1\linewidth]{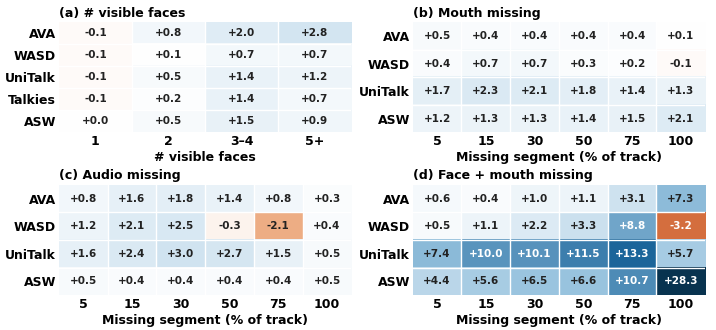}}
\vspace{-1.1em}
\caption{Inter-face context gain
($\Delta\mathrm{mAP} =
\mathrm{mAP}_{\text{ROAM-ASD-C}} -
\mathrm{mAP}_{\text{ROAM-ASD}}$)
with (a) different numbers of visible faces and under continuous
(b) mouth, (c) audio, and (d) all-visual missing segments.}
\label{fig:mix}
\vspace{-1.5em}
\end{figure}
\begin{table}[!htbp]
\vspace{-1.5em}
\centering
\caption{mAP ($\uparrow$, \%) on AVA with different numbers of visible faces.}
\vspace{0.2em}
\label{tab:ava_num_faces}
\resizebox{0.29\textwidth}{!}{
\setlength{\tabcolsep}{4pt}
\begin{tabular}{lccc}
\toprule
Method & 1 face & 2 faces & 3 faces \\
\midrule
ASC~\cite{alcazar2020active}
    & $91.8$ & $83.8$ & $67.6$ \\
MAAS~\cite{alcazar2021maas}
    & $93.3$ & $85.8$ & $68.2$ \\
TalkNet~\cite{tao2021someone}
    & $95.4$ & $89.6$ & $80.3$ \\
ASDNet~\cite{kopuklu2021design}
    & $95.7$ & $92.4$ & $83.7$ \\
EASEE~\cite{alcazar2022end}
    & $96.5$ & $92.4$ & $83.9$ \\
Light-ASD~\cite{liao2023light}
    & $96.2$ & $92.6$ & $84.4$ \\
MuSED~\cite{tao2024enhancing}
    & $96.7$ & $93.6$ & $87.8$ \\
LoCoNet~\cite{wang2024loconet}
    & $97.0$ & $94.6$ & $87.4$ \\
\midrule
ROAM-ASD~\cite{wang2026roam}
    & $\mathbf{97.9}$ & $95.8$ & $89.2$ \\
\textbf{ROAM-ASD-C}
    & $97.8$ & $\mathbf{96.6}$ & $\mathbf{91.4}$ \\
\bottomrule
\end{tabular}
}
\vspace{-1em}
\end{table}

Table~\ref{tab:overall} first compares ROAM-ASD-C with ROAM-ASD and prior ASD systems. Adding inter-face contextual reasoning consistently improves ROAM-ASD across all five benchmarks, reaching $96.8\%$, $98.9\%$, $88.8\%$, $98.7\%$, and $99.5\%$ mAP on AVA, WASD, UniTalk, Talkies, and ASW, respectively. We next examine where these gains arise. Fig.~\ref{fig:wasdunitalk}
breaks down the improvement across the individual WASD and UniTalk conditions. On the relatively easy WASD conditions, \emph{OC} and \emph{SI}, ROAM-ASD already achieves $99.6\%$ and $99.7\%$ mAP, leaving little room for further improvement. In contrast, under the more challenging \emph{HVN} and \emph{SS} conditions, where ROAM-ASD obtains $96.3\%$ and $96.2\%$ mAP, ROAM-ASD-C improves by $0.4$ and $1.5$ mAP, respectively. A similar pattern is observed on UniTalk, where contextual reasoning improves all four conditions, with the largest gains under \emph{Noise} and \emph{Crowded}, both by $1.5$ mAP. These results suggest that inter-face context becomes more useful when the evidence available from an individual face track is less reliable. This effect is also reflected in the number of simultaneously visible faces. As shown in Table~\ref{tab:ava_num_faces}, ROAM-ASD-C performs essentially the same as ROAM-ASD with only one visible face, while the gain increases to $0.8$ mAP with two faces and $2.2$ mAP with three faces. The same tendency is observed across the five benchmarks in Fig.~\ref{fig:mix}(a). Examples of crowded and competing-speaker scenarios are available at\url{https://wangpuup.github.io/ROAM-ASD-C/crowds.html}. 
 
Given that contextual gains increase when a candidate's own evidence becomes less reliable, we next evaluate whether other visible candidates can compensate for missing target-track observations.
Fig.~\ref{fig:mix}(d) shows that ROAM-ASD-C provides substantial gains under long all-visual missing segments (face and mouth). At $75\%$ removal, ROAM-ASD-C improves over ROAM-ASD by $3.1$, $8.8$, $13.3$, and $10.7$ mAP on AVA, WASD, UniTalk, and ASW, respectively. 
Examples illustrating this behavior are available at\url{https://wangpuup.github.io/ROAM-ASD-C/modality.html}.

Finally, we evaluate whether speaker registration can extend ASD beyond the currently visible candidates. By design, conventional face-level ASD, including ROAM-ASD-C, can only make predictions for tracked visible faces. When the visual streams are fully removed, scene audio alone does not provide the identity information required to determine which candidate produced the speech. Speaker registration provides such identity information through the enrollment representation and therefore enables speaker-conditioned activity detection even without visible evidence. As shown in Table~\ref{tab:registration}, under complete visual absence, the registered-speaker branch retains $80.4\%$, $78.0\%$, $81.8\%$, and $60.3\%$ AP on ASW, AVA, WASD, and UniTalk, respectively, substantially higher than the corresponding face-ASD performance of $52.8\%$, $58.9\%$, $42.4\%$, and $36.4\%$. Importantly, this additional capability does not affect standard face-level ASD: Table~\ref{tab:overall} shows that ROAM-ASD-CR preserves the performance of ROAM-ASD-C. The model also remains robust to over-registration, with FA rates for out-of-scene registrants remaining at or below $1.3\%$ even when $20$ additional speakers are enrolled. Together, these results show that speaker registration provides a complementary way to reason about speakers beyond the currently visible scene while preserving conventional face-level ASD performance.
\begin{table}[!htbp]
\centering
\vspace{-1.5em}
\caption{Speaker-registration performance of ROAM-ASD-CR (\%).}
\vspace{0.2em}
\label{tab:registration}
\resizebox{0.49\textwidth}{!}{
\begin{tabular}{l|ccc|cc}
\toprule
& \multicolumn{2}{c}{Registered-speaker AP}
& Face ASD & Association & FA \\
\cmidrule(lr){2-3}
\cmidrule(lr){4-4}
\cmidrule(lr){5-5}
\cmidrule(lr){6-6}
Dataset & Full & Visual absent & Visual absent & Acc. & Out-of-scene $+20$ \\
\midrule
ASW     & $93.3$ & $80.4$ & $52.8$ & $98.4$ & $0.0$ \\
AVA     & $89.0$ & $78.0$ & $58.9$ & $95.5$ & $0.04$ \\
WASD    & $87.4$ & $81.8$ & $42.4$ & $92.7$ & $1.3$ \\
UniTalk & $72.4$ & $60.3$ & $36.4$ & $79.9$ & $0.0$ \\
\bottomrule
\end{tabular}
}
\vspace{-1.5em}
\end{table}
\section{Conclusion}
We presented ROAM-ASD-C and ROAM-ASD-CR, extending ROAM-ASD with
inter-face contextual reasoning and speaker registration. ROAM-ASD-C consistently improves face-level ASD, with larger gains in crowded and visually degraded conditions. ROAM-ASD-CR further enables registered-speaker activity detection when visual evidence is unavailable while preserving face-level ASD performance. Together, the two extensions broaden ASD from independent visible face tracks
toward contextual and identity-aware reasoning.

\newpage
\section*{Acknowledgment}
The authors thank Xiaomi Corporation, Beijing, China for their support in pursuing this research.
\bibliographystyle{IEEEbib}
\bibliography{strings,refs}

@article{wang2026roam,
  title={ROAM-ASD: Robust Open-World Active Speaker Detection with Flexible Multimodal Fusion},
  author={Wang, Pu and Van hamme, Hugo},
  journal={arXiv preprint arXiv:2609.26648},
  year={2026}
}

@inproceedings{jung2024talknce,
  title={Talknce: Improving active speaker detection with talk-aware contrastive learning},
  author={Jung, Chaeyoung and Lee, Suyeon and Nam, Kihyun and Rho, Kyeongha and Kim, You Jin and Jang, Youngjoon and Chung, Joon Son},
  booktitle={ICASSP 2024-2024 IEEE International Conference on Acoustics, Speech and Signal Processing (ICASSP)},
  pages={8391--8395},
  year={2024},
  organization={IEEE}
}

@inproceedings{wang2024loconet,
  title={Loconet: Long-short context network for active speaker detection},
  author={Wang, Xizi and Cheng, Feng and Bertasius, Gedas},
  booktitle={2024 IEEE/CVF Conference on Computer Vision and Pattern Recognition (CVPR)},
  pages={18462--18472},
  year={2024},
  organization={IEEE}
}

@article{tao2024enhancing,
  title={Enhancing real-world active speaker detection with multi-modal extraction pre-training},
  author={Tao, Ruijie and Qian, Xinyuan and Das, Rohan Kumar and Gao, Xiaoxue and Wang, Jiadong and Li, Haizhou},
  journal={IEEE Transactions on Multimedia},
  volume={27},
  pages={2362--2373},
  year={2024},
  publisher={IEEE}
}

@inproceedings{liao2023light,
  title={A light weight model for active speaker detection},
  author={Liao, Junhua and Duan, Haihan and Feng, Kanghui and Zhao, Wanbing and Yang, Yanbing and Chen, Liangyin},
  booktitle={2023 IEEE/CVF Conference on Computer Vision and Pattern Recognition (CVPR)},
  pages={22932--22941},
  year={2023},
  organization={IEEE}
}

@article{roxo2024wasd,
  title={WASD: A wilder active speaker detection dataset},
  author={Roxo, Tiago and Costa, Joana Cabral and In{\'a}cio, Pedro RM and Proen{\c{c}}a, Hugo},
  journal={IEEE Transactions on Biometrics, Behavior, and Identity Science},
  volume={7},
  number={1},
  pages={61--70},
  year={2024},
  publisher={IEEE}
}

@inproceedings{nguyen2025revisiting,
  title={Revisiting Active Speaker Detection: An In-the-Wild Benchmark for Generalization and Robustness},
  author={Nguyen, Le Thien Phuc and Yu, Zhuoran and Cao, Khoa Quang Nhat and Guo, Yuwei and Pham, Tu Ho Manh and Nguyen, Tuan Tai and Vo, Toan Ngo Duc and Poon, Lucas and Nguyen, Tuan Khai and Lee, Soochahn and others},
  booktitle = {{Interspeech 2026}},
  year={2026}
}

@inproceedings{alcazar2020active,
  title={Active speakers in context},
  author={Alc{\'a}zar, Juan Le{\'o}n and Caba, Fabian and Mai, Long and Perazzi, Federico and Lee, Joon-Young and Arbel{\'a}ez, Pablo and Ghanem, Bernard},
  booktitle={2020 IEEE/CVF Conference on Computer Vision and Pattern Recognition (CVPR)},
  pages={12462--12471},
  year={2020},
  organization={IEEE}
}

@inproceedings{alcazar2021maas,
  title={Maas: Multi-modal assignation for active speaker detection},
  author={Alc{\'a}zar, Juan Le{\'o}n and Caba, Fabian and Thabet, Ali K and Ghanem, Bernard},
  booktitle={2021 IEEE/CVF International Conference on Computer Vision (ICCV)},
  pages={265--274},
  year={2021},
  organization={IEEE}
}

@inproceedings{tao2021someone,
  title={Is someone speaking? exploring long-term temporal features for audio-visual active speaker detection},
  author={Tao, Ruijie and Pan, Zexu and Das, Rohan Kumar and Qian, Xinyuan and Shou, Mike Zheng and Li, Haizhou},
  booktitle={Proceedings of the 29th ACM international conference on multimedia},
  pages={3927--3935},
  year={2021}
}

@inproceedings{kopuklu2021design,
  title={How to design a three-stage architecture for audio-visual active speaker detection in the wild},
  author={K{\"o}p{\"u}kl{\"u}, Okan and Taseska, Maja and Rigoll, Gerhard},
  booktitle={2021 IEEE/CVF International Conference on Computer Vision (ICCV)},
  pages={1173--1183},
  year={2021},
  organization={IEEE}
}

@inproceedings{alcazar2022end,
  title={End-to-end active speaker detection},
  author={Alcazar, Juan Leon and Cordes, Moritz and Zhao, Chen and Ghanem, Bernard},
  booktitle={European Conference on Computer Vision},
  pages={126--143},
  year={2022},
  organization={Springer}
}

@inproceedings{roth2020ava,
  title={Ava active speaker: An audio-visual dataset for active speaker detection},
  author={Roth, Joseph and Chaudhuri, Sourish and Klejch, Ondrej and Marvin, Radhika and Gallagher, Andrew and Kaver, Liat and Ramaswamy, Sharadh and Stopczynski, Arkadiusz and Schmid, Cordelia and Xi, Zhonghua and others},
  booktitle={ICASSP 2020-2020 IEEE international conference on acoustics, speech and signal processing (ICASSP)},
  pages={4492--4496},
  year={2020},
  organization={IEEE}
}

@inproceedings{kim21k_interspeech,
  title     = {{Look Who’s Talking: Active Speaker Detection in the Wild}},
  author    = {You Jin Kim and Hee-Soo Heo and Soyeon Choe and Soo-Whan Chung and Yoohwan Kwon and Bong-Jin Lee and Youngki Kwon and Joon Son Chung},
  year      = {2021},
  booktitle = {{Interspeech 2021}},
  pages     = {3675--3679},
  doi       = {10.21437/Interspeech.2021-2041},
  issn      = {2958-1796},
}

@inproceedings{jiang23c_interspeech,
  title     = {{Target Active Speaker Detection with Audio-visual Cues}},
  author    = {Yidi Jiang and Ruijie Tao and Zexu Pan and Haizhou Li},
  year      = {2023},
  booktitle = {{Interspeech 2023}},
  pages     = {3152--3156},
  doi       = {10.21437/Interspeech.2023-574},
  issn      = {2958-1796},
}

\end{document}